\documentclass{article}
\usepackage[utf8]{inputenc}
\usepackage[T1]{fontenc}
\usepackage{braket}
\usepackage{listings}
\usepackage{xcolor}
\usepackage{blindtext}
\usepackage{graphicx}
\usepackage{authblk}
\usepackage{float}
\usepackage{appendix}
\usepackage{hyperref}
\usepackage{lscape}
\usepackage{utfsym}
\usepackage{amsmath}
\usepackage{multirow}
\usepackage{array}
\usepackage{helvet}
\usepackage{tikz}
\usepackage{soul}
\usepackage{subfig}
\usepackage{pgfplots}
\usepackage{colortbl}

\usepgfplotslibrary{groupplots}

\title{A pragma based C++ framework for hybrid quantum/classical computation}

\author[1]{Arnaud GAZDA}
\author[1,2]{Océane KOSKA}

\affil[1]{Eviden Quantum Lab, Les Clayes-sous-Bois, France}
\affil[2]{Laboratoire Interdisciplinaire des Sciences Numériques, Gif-sur-Yvette, Saclay, France}

\newcolumntype{?}{!{\vrule width 2pt}}

\newenvironment{explaination}{
    \setlength{\leftskip}{5mm}

}{

    \setlength{\leftskip}{0mm}
}

\definecolor{colorstring}{rgb}{0.69, 0.37, 0}
\definecolor{colortype}{rgb}{0.07, 0.57, 0.05}
\definecolor{colorpragma}{rgb}{0.7, 0.05, 0.8}
\definecolor{colorcomment}{rgb}{0.21, 0.55, 0.8}
\definecolor{colorbg}{rgb}{0.95,0.95,0.92}
\definecolor{colornonterm}{rgb}{0.2,0.5,1}
\definecolor{colornontermbutterm}{rgb}{0.4,0.6,0.7}
\definecolor{colorterm}{rgb}{0.5,0.5,0.5}

\lstdefinestyle{desert_scheme}{
    backgroundcolor=\color{colorbg},
    commentstyle=\color{colorcomment},
    keywordstyle=\color{colortype},
    stringstyle=\color{colorstring},
    basicstyle=\ttfamily\small,
    escapechar={|},
    breakatwhitespace=false,
    frame=single,
    breaklines=true,
    captionpos=b,
    keepspaces=true,
    numbers=left,
    numbersep=5pt,
    showspaces=false,
    showstringspaces=false,
    showtabs=false,
    tabsize=4
}

\newcommand{\pragma}[1]{\textbf{\ttfamily\small\color{colorpragma}\#pragma #1}}
\newcommand{\qbool}[0]{{\ttfamily\small\color{colortype}qbool }}
\newcommand{\castqbool}[0]{\ttfamily\small({\color{colortype}qbool}) }
\newcommand{\double}[0]{{\ttfamily\small\color{colortype}double }}
\newcommand{\qint}[1]{{\ttfamily\small\color{colortype}qint\_t}<#1> }
\newcommand{\quint}[1]{{\ttfamily\small\color{colortype}quint\_t}<#1> }
\newcommand{\qinteight}[0]{{\ttfamily\small\color{colortype}qint8\_t }}
\newcommand{\quinteight}[0]{{\ttfamily\small\color{colortype}quint8\_t }}
\newcommand{\cuint}[1]{{\ttfamily\small\color{colortype}uint#1\_t }}
\newcommand{\constexpr}[0]{{\ttfamily\small\color{colortype}constexpr }}
\newcommand{\nonterm}[1]{$\langle${\color{colornonterm}#1}$\rangle$}
\newcommand{\term}[1]{{\color{colorterm}``#1''}}
\newcommand{\nontermbutterm}[1]{{\color{colornontermbutterm}$\langle$#1$\rangle$}}
\newcommand{\namedparagraph}[1]{\paragraph{#1}\mbox{}\\}
\newcommand{\checkmark}[0]{\small\usym{1F5F8}}
\newcommand{\unclear}[0]{$\boldsymbol{\sim}$}
\newcommand{\best}{\cellcolor{green!25}}

\definecolor{torange}{RGB}{255,182,161}
\definecolor{tgray}{RGB}{127,150,157}
\definecolor{tblue}{RGB}{127,192,214}

\newcommand{\tikzfont}{\fontfamily{phv}\selectfont \Large}
\newcommand{\device}[5]{
    \draw[dashed] (#1,#2) rectangle ++(5,-4.5) node[pos=0,anchor=north west] {\tikzfont #3};
    \draw[fill=torange,torange] (#1 + 0.3,#2 - 0.75) rectangle ++(4.4,-0.95) node[pos=.5,black] {\tikzfont Controller};
    \draw[fill=#5,#5] (#1 + 0.3,#2 - 1.85) rectangle ++(4.4,-2.35) node[pos=.5,black] {\tikzfont #4};
    \draw[<->,very thick] (#1 + 2.5,#2 - 0.5) -- ++(0,1.25);
}

\providecommand{\keywords}[1]
{
    \small	
    \textbf{\textit{Keywords---}} #1
}

\newenvironment{code}{
    \medskip

    \begin{center}
    \begin{minipage}{0.85 \textwidth}
}{
    \end{minipage}
    \end{center}
}

\newcommand{\grammar}[1]{
    \begin{center}
    \scalebox{0.9}{\begin{tabular}{lll}
        #1
    \end{tabular}}
    \end{center}
}

\begin{document}


\maketitle

\begin{abstract}
Quantum computers promise exponential speed ups over classical computers for various tasks. This emerging technology is expected to have its first huge impact in High Performance Computing (HPC), as it can solve problems beyond the reach of HPC. To that end, HPC will require quantum accelerators, which will enable applications to run on both classical and quantum devices, via hybrid quantum-classical nodes. Hybrid quantum-HPC applications should be scalable, executable on Quantum Error Corrected (QEC) devices, and could use quantum-classical primitives. However, the lack of scalability, poor performances, and inability to insert classical schemes within quantum applications has prevented current quantum frameworks from being adopted by the HPC community.

This paper specifies the requirements of a hybrid quantum-classical framework compatible with HPC environments, and introduces a novel hardware-agnostic framework called Q-Pragma. This framework extends the classical programming language C++ heavily used in HPC via the addition of pragma directives to manage quantum computations.
\end{abstract}




\section{Introduction}

Quantum computing is a paradigm of computer science that uses quantum physics to solve some computationally hard problems faster than a classical computer. These problems, such as cracking encryption \cite{Shor_1997} or simulating quantum chemistry systems \cite{quantum_chemistry,grid_based_quantum_chemistry}, are currently solved using High Performance Computing (HPC) systems, but can be tackled more efficiently using quantum algorithms, suggesting that quantum computing will revolutionize HPC. However, these quantum algorithms are not expected to replace existing classical applications; they will be integrated inside existing softwares to avoid some heavy classical computations. For example, the application that cracks encryption will use a quantum processor to find an integer divisor while parsing a SSL certificate or decoding an encrypted message will require classical resources. Applications using quantum resources will be hybrid, requiring both quantum and classical resources. Moreover, the classical code used in these hybrid applications is likely already implemented. Then, extending classical applications to integrate quantum kernels is a crucial step in the process of making quantum devices useful.

\vspace{5mm}

This kind of integration has already been done in classical computing, using more than one type of processor or core. For example, in general-purpose computing on Graphical Processing Units (GPU), the Central Processing Unit (CPU) delegates some tasks to the GPU acting as an accelerator \cite{gpgpu}. The emergence of new programming tools such as CUDA \cite{cuda} and OpenMP \cite{openmp, openmp_gpu} have made classical hybridization possible, based on a well-defined hardware connection between the GPU and the CPU. Similarly, tasks could be offloaded to a Quantum Processing Unit (QPU), by taking into account the interaction between classical and quantum hardware. The main challenge would be to connect the quantum computing systems with classical ones, and to define the programming interface to communicate with such quantum resources.

\vspace{5mm}

Proposition of architectures integrating quantum devices into a HPC node\footnote{A node is the minimal entity on which an Operating System (OS) can be installed} were recently published \cite{quantum_computer_architecture, quantum_computer_for_hpc} (these nodes are referred to as ``hybrid quantum-classical nodes''). In these designs, the classical resource (named \textit{Host}) is placed close to the quantum device to reduce latency and to enhance the quantum computation efficiency. Reducing the latency improves the computation and allows for efficient data-transfers between the components of the node. In these designs, the Host is directly connected to a controller, which is a piece of classical software and electronics responsible for manipulating the quantum state within the quantum device.

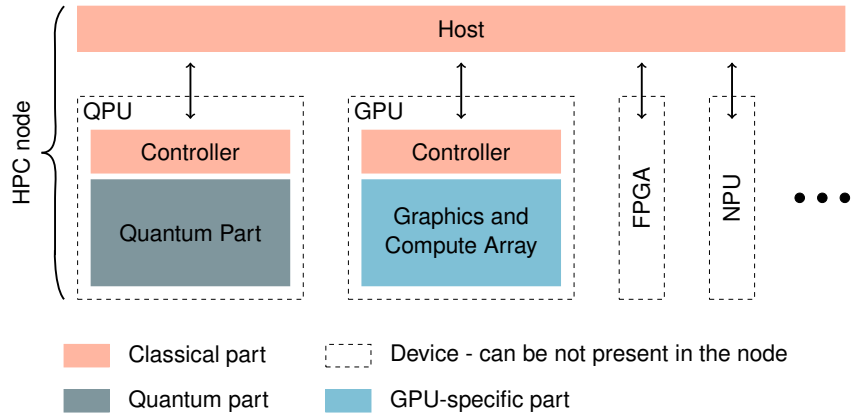
\begin{figure}[!h]
    \centering
    \scalebox{0.6}{\begin{tikzpicture}

\node[rotate=90] at (-1.25,-2.25) {\tikzfont HPC node};
\draw[very thick] (-0.3,1)    to[out=230,in=90] (-0.5,0)    -- (-0.5,-1.25) to[out=-90,in=0] (-0.8, -2.25);
\draw[very thick] (-0.3,-5.5) to[out=130,in=90] (-0.5,-4.5) -- (-0.5,-3.25) to[out=90,in=0]  (-0.8, -2.25);

\draw[fill=torange,torange] (0,0) rectangle (17,1) node[pos=.5,black] {\tikzfont Host};

\device{0}{-1}{QPU}{Quantum Part}{tgray}

\device{6}{-1}{GPU}{\begin{tabular}{c} Graphics and \\ Compute Array \end{tabular}}{tblue}

\draw[dashed] (12, -1) rectangle ++(1,-4.5) node[pos=0.5, rotate=90] {\tikzfont FPGA};
\draw[<->,very thick] (12.5,-1.5) -- ++(0,1.25);

\draw[dashed] (14, -1) rectangle ++(1,-4.5) node[pos=0.5, rotate=90] {\tikzfont NPU};
\draw[<->,very thick] (14.5,-1.5) -- ++(0,1.25);

\draw[fill=black] (16,-3.25) circle (3pt);
\draw[fill=black] (16.5,-3.25) circle (3pt);
\draw[fill=black] (17,-3.25) circle (3pt);

\draw[fill=torange,torange] (-0.3,-6.5) rectangle ++(1,-0.5);
\node[anchor=west] at (1, -6.75) {\tikzfont Classical part};

\draw[fill=tgray,tgray] (-0.3,-7.5) rectangle ++(1,-0.5);
\node[anchor=west] at (1, -7.75) {\tikzfont Quantum part};

\draw[dashed] (5.5,-6.5) rectangle ++(1,-0.5);
\node[anchor=west] at (6.8, -6.75) {\tikzfont Device - can be not present in the node};

\draw[fill=tblue,tblue] (5.5,-7.5) rectangle ++(1,-0.5);
\node[anchor=west] at (6.8, -7.75) {\tikzfont GPU-specific part};

\end{tikzpicture}}
    \caption{Hybrid HPC node - composed of a Host and one or several devices. The QPU is connected to the Host like any other device.}
    \label{fig:node-design}
\end{figure}

In Figure \ref{fig:node-design}, the QPU controller receives instructions from the Host to execute them on the quantum part. It schedules these instructions, and can execute classical computations that interact with the quantum part. Incorporating classical computations alongside quantum computation is used to manage Quantum Error Correction (QEC), but can also be used to optimize a quantum routine. For example, this method is used to reduce the gate count of arithmetic operations \cite{WindowedArith}. The incorporation of classical computation directly impacts the design of a hybrid quantum-HPC framework.

\vspace{5mm}

Existing quantum frameworks (like OpenQASM 3 \cite{openqasm3}, Q\# \cite{Qsharp}, and Scaffold \cite{scaffold}) do not take advantage of this hardware integration, by restricting CPU/QPU interactions. Moreover, running a quantum routine within an existing application appears challenging, as the discussion on the interoperability between HPC and quantum computing remains limited. Then, redesigning classical-quantum programming is a foremost requirement, but this redesign can also simplify the way hybrid applications are implemented. To do so, defining the specifications of a hybrid quantum-HPC framework is necessary.

\vspace{5mm}

This paper proposes a definition of a hybrid quantum-HPC framework based on 7 criteria, inspired by well-established classical and quantum hybrid frameworks. Based on these criteria, a new C++ quantum-HPC framework called ``Q-Pragma'' is introduced. This framework integrates modern programming concepts to satisfy HPC constraints and to simplify the integration of quantum kernels in existing applications. Nevertheless, these constraints are not an obstacle to use Q-Pragma in a non-HPC context, and can even simplify the development of hybrid algorithms. For instance, advanced quantum algorithms can be written using only few lines of code (see Appendix \ref{app:examples}). A core concept of Q-Pragma is to use pragma directives to add quantum-classical hybridization capabilities to C++. Q-Pragma is a hardware-agnostic framework built on existing HPC design patterns, making it portable, scalable, and usable with classical HPC frameworks.

In the next section, the term ``hybrid quantum-HPC'' will be used, since this section focus on the integration of a quantum device in a HPC system. Subsequently, the term ``hybrid quantum-classical'' will be preferred to discuss about hybridization in general (and not only HPC).

\newpage


\section{Defining a hybrid quantum-HPC framework}

Developing a hybrid quantum-HPC framework seamlessly interoperable with HPC languages is of paramount importance. Such interoperability would guarantee compatibility and facilitate smooth integration with current HPC infrastructures and software ecosystem. Additionally, these specifications play a critical role in shaping the framework functionalities and performance.

\vspace{5mm}

HPC is already hybrid, as many existing applications rely on GPUs. This section focus on the existing classical hybrid frameworks, to highlight HPC-friendly design patterns which can be used in a quantum-HPC framework. Then, this section focus on existing quantum frameworks to identify features that need to be implemented in quantum-HPC frameworks.


\subsection{Classical hybrid computing}

Within the realm of HPC, the most prominent languages are Fortran \cite{fortran}, C \cite{C}, and C++ \cite{C++}. Fortran, the oldest among them, maintains its active presence in HPC, primarily due to its utilization in legacy codes. Meanwhile, C stands out for its precise memory management and system operations, making it particularly powerful for performance optimization. Additionally, a vast range of libraries and tools like BLAS \cite{BLAS} and LAPACK \cite{lapack} are available for C, enriching its development ecosystem. On the other hand, despite its greater complexity compared to C and Fortran, C++ offers the advantages of object-oriented programming, significantly enhancing code maintainability in large-scale HPC applications. Furthermore, C++ has template metaprogramming capabilities that enable compile-time code generation, thus enhancing performance further. These three programming languages possess interoperability, meaning that a function written in any of these languages can be invoked or utilized within another language. Creating a framework that works easily with today's HPC languages allows us to use pre-existing tools including parallel programming tools and optimized libraries.

In classical hybrid computing, interoperability with one of these programming languages prevails. There are two types of classical hybrid computing frameworks, frameworks based on a new programming language, and frameworks extending an existing HPC language. 

\namedparagraph{Hybrid frameworks based on a new programming language}
Some hybrid applications already use third-party languages to run code on specific devices. For example, CUDA \cite{cuda} and OpenCL \cite{openCL} are programming languages designed to take advantage of GPUs.

CUDA (Compute Unified Device Architecture) is a parallel computing platform and programming model developed by NVIDIA, specifically designed to exploit the computational power of GPUs for general-purpose computing tasks. CUDA provides a C-like programming interface, allowing developers to express GPU computations within their existing C or C++ codebase.

Similarly OpenCL is an open standard parallel programming framework that targets heterogeneous computing environments like CPUs, GPUs and FGPAs. OpenCL also supports several programming languages such as C, C++ and Python.

This design pattern provides the concept of \textit{kernel} (i.e. a device-specific function which can be called from a HPC language) and provides tools to allocate memory on the device. This design pattern has been chosen by some quantum frameworks, including CUDA Quantum \cite{cudaquantum}, Q\# \cite{Qsharp}, OpenQASM 3 \cite{openqasm3}, and Scaffold \cite{scaffcc}. Howeover, with the exception of CUDA Quantum, none of these programming languages incorporate the concept of kernel, preventing these frameworks from being used in HPC environments.

\namedparagraph{Hybrid frameworks extending existing programming languages}
Another option is to enhance an existing language to streamline integration. Among conventional hybrid programming solutions, OpenMP \cite{openmp, openmp_gpu} (Open Multi-Processing) stands out as a prominent programming model engineered to seamlessly integrate GPU-specific code within established programming languages like C, C++, or Fortran. OpenMP is based on a collection of compiler directives, library functions, and environment variables. These directives, highlighted by pragmas, allow programmers to define code sections for GPU execution without having to learn a new programming language. They serve the purpose of adding device specific instructions as well as specifying the code and memory locality (i.e. selecting the device on which a function is executed, or a variable stored). This design pattern has been chosen by QCOR \cite{qcor}, by adding function attribute specifiers to the C++ language.

\vspace{5mm}

These two classical design patterns address a critical concern of classical hybrid environments, namely code and memory locality control. This concern should be addressed by quantum-HPC frameworks through the following specification:

\namedparagraph{Code and memory locality}
\textit{A quantum-HPC framework provides tools to control code and memory locality.}

\vspace{5mm}

Existing classical HPC frameworks are extended with device-specific features to encompass the full spectrum of capabilities offered by the device. Thus, the capabilities of a QPU must be defined in order to build a quantum-HPC framework.


\subsection{Quantum computing features for HPC}

Quantum computing will impact HPC. To harness this acceleration, quantum computing features for HPC must be listed. A set of hardware capabilities, through the seven DiVicenzo's criteria \cite{DiVincenzo2000}, has defined what a universal quantum computer is. This section defines a list of software capabilities to better take advantage of a such device. An overview of these specific capabilities has never been done before. Inspired by the state of the art, this subsection outlines the main features of a quantum-HPC framework. It involves using millions of physical qubits, correcting errors for accurate results, and spreading computations across multiple QPUs for better processing. These requirements will need specific capabilities from the QPU.

\paragraph{Dynamic interaction}
\textit{Classical memory or quantum memory can be allocated on a QPU and manipulated from the Host, at runtime.}

\vspace{2mm}

Hybrid quantum-HPC applications require their classical part to interact with their quantum part. To interact with a quantum computation, the Host needs to send and receive information from the QPU. The Host should then be able to access QPU's memory, by reading and writing data directly on the QPU.

\paragraph{Scalability}
\textit{Hybrid quantum-HPC frameworks should support algorithms with arbitrary big number of qubits, or instructions.}

\vspace{2mm}

Pratical quantum-HPC applications, like Shor algorithm running on an error-corrected QPU \cite{Shor_8h}, would require millions of physical qubits (the \textit{scalability} is a hardware requirement, since it is a DiVicenzo's criteria). Then, any hybrid quantum-HPC framework must be able to handle such large amount of physical qubits. Nowadays, the circuit formalism used to describe quantum algorithms represents, on a timeline, the different operations that will be applied on the quantum state. This formalism is static and is used by most of today's quantum frameworks like myQLM \cite{myqlm}, Qiskit \cite{qiskit}, and Cirq \cite{cirq}. Nevertheless, the circuit formalism is not designed to represent billions of operations acting on millions of qubits, as it would be necessary to create a data structure of several gigabytes to represent a single instance of the problem. Moreover, the integration of classical operations that have an impact on the quantum computation is not compatible with this formalism.

\paragraph{Typing}
\textit{Quantum registers are typed to simplify the manipulation of huge structures.}

\vspace{2mm}

Today, the majority of classical programming languages utilizes typing, where source code handles not just booleans but also more complex data types like integers, floating points, and arrays. Employing typing in a language streamlines the source code and empowers users to create sophisticated algorithms. The use of typing in a framework is essential to ensure scalability. A quantum-HPC framework should offer predefined quantum data types while also permitting users to define their own custom quantum structures.

\paragraph{Reversibility}
\textit{Pure quantum operations are reversible.}

\vspace{2mm}

Quantum reversibility is a fundamental concept in quantum computing, referring to the ability to reverse quantum operations and undo their effects \cite{nielsen_chuang}. Any quantum routine written in a hybrid framework should be reversible.

\paragraph{Controllability}
\textit{Pure quantum operations are controllable.}

\vspace{2mm}

In quantum computing, control gates such as Toffoli gates have a crucial function as they facilitate the creation of entanglement \cite{nielsen_chuang}. This entanglement, along with the concept of superposition, empowers quantum computers to execute advanced computations that greatly surpass the classical computer capabilities. The ability to control a quantum routine constitutes a fundamental concept within a quantum-HPC framework.

\paragraph{Safe uncomputation}
\textit{Quantum registers can be reset to $\ket{0}$ state without any measurement.}

\vspace{2mm}

Unlike classical computing, resetting a qubit is dangerous (reset being composed of a measurement, and optionally a X gate). A qubit may be entangled with another one, so resetting a qubit could have an impact on other qubits. Safe uncomputation is a concept defined in existing quantum frameworks like Quipper \cite{quipper} or Silq \cite{silq}, relying on the reversibility of quantum unitary operations to reset a qubit to the $\ket{0}$ state.

\vspace{5mm}

None of the existing quantum frameworks fulfill the aforementioned requirements, see Table \ref{tab:frameworks}. A new C++ framework for hybrid quantum-classical computing fulfilling all these requirements is introduced. This framework, compatible with HPC environments, is composed of both a C++ library, as well as a compiler plugin providing C++ pragmas to control the quantum computation.

\begin{landscape}
\begin{table}

\begin{tabular}{cl}
\checkmark & Fully supported \\
\unclear   & Partially supported \\
\end{tabular}

\vspace{10mm}

\centering
\scalebox{0.9}{
\setlength\doublerulesep{5pt}
\begin{tabular}{|l?c|c|c|c|c|c|c|} 
 \hline
 Framework & Locality & Dynamic interaction & Scalability  & Typing & Reversibility & Controllability & Safe uncomputation \\ 
 \hline
 \hline
 Cirq \cite{cirq} & & & & & \checkmark & \checkmark & \\
 \hline
 CUDA Quantum \cite{cudaquantum} & \unclear & & & & \checkmark & \checkmark & \unclear \\
 \hline
 myQLM \cite{myqlm} & & & & \checkmark & \checkmark & \checkmark & \unclear \\
 \hline
 OpenQASM 3 \cite{openqasm3} & & \checkmark & \checkmark & & \checkmark & \checkmark & \\
 \hline
 ProjectQ \cite{ProjectQ_1, ProjectQ_2} & & \checkmark & \checkmark & & \checkmark & \checkmark & \unclear \\
 \hline
 Q\# \cite{Qsharp} & & \checkmark & \unclear & & \checkmark & \checkmark & \unclear \\
 \hline
 QCOR \cite{qcor} & \unclear & & & & \checkmark & \checkmark & \\
 \hline
 Qiskit \cite{qiskit} & & & & & \checkmark & \checkmark & \\
 \hline
 Quipper \cite{quipper} & & & & & \checkmark & \checkmark & \checkmark \\
 \hline
 Scaffold \cite{scaffold} & & & & & & \checkmark & \\
 \hline
 Silq \cite{silq} & & & & \checkmark & \checkmark & \checkmark & \checkmark \\
 \hline
 \multicolumn{1}{c}{\phantom{x}} \vspace{-2\tabcolsep}\vspace{\doublerulesep} \\
 \cline{1-5}
 CUDA \cite{cuda} & \checkmark & \checkmark & \checkmark & \checkmark \\ 
 \cline{1-5}
 OpenCL \cite{openCL} & \checkmark & \checkmark & \checkmark & \checkmark \\ 
 \cline{1-5}
 OpenMP \cite{openmp, openmp_gpu} & \checkmark & \checkmark & \checkmark & \checkmark \\ 
 \cline{1-5}
\end{tabular}}

\caption{Features implemented by existing hybrid frameworks - the first group corresponds to quantum frameworks, the second one corresponds to classical frameworks (see Appendix \ref{app:features} for details).}
\label{tab:frameworks}
\end{table}
\end{landscape}

\newpage


\section{Q-Pragma: a C++ hybrid framework}

This section introduces a new framework, called ``Q-Pragma'' that bridges the classical and quantum computing worlds, even within HPC ecosystems. Q-Pragma is a C++-based framework that not only empowers classical applications with quantum capabilities but also satisfies the \textit{code and memory locality} requirement. This framework constitutes a first step towards the development of hybrid quantum-classical applications, and adds the power of quantum computing to the computational efficiency and scalability offered by HPC systems.

At its core, this framework comprises a versatile C++ library designed to facilitate the seamless integration of quantum operations into classical C++ code. This integration enables developers to explore quantum algorithms and harness quantum hardware resources without completely overhauling their existing classical codebases. Moreover, C++ pragmas specifically tailored to extend the language expressive power are introduced, allowing for the efficient description of quantum operations within classical code, thus forming a bridge between the two computing paradigms.

\vspace{5mm}

In Q-Pragma, the \textit{dynamic interaction} requirement is satisfied through the memory sharing enabled by the HPC link between Host and QPU \cite{quantum_computer_for_hpc}. This is discussed in Subsection \ref{sec:locality}.

Additionally, the QPU connected to the Host is directly targeted by Q-Pragma, simplifying hybrid application development. This targeting relies on a C++ library available on the node. This library provides a connection with the QPU, and instructions are directly streamed through this library. This fulfills the \textit{scalability} requirement. Streaming instructions avoid the drawbacks of the circuit structure, permitting the framework to manipulate large scale quantum computations, as well as allowing classical computation to interact with the quantum part. Streaming instructions is a known pattern used by OpenQASM 3 \cite{openqasm3} or ProjectQ \cite{ProjectQ_1, ProjectQ_2}.

\vspace{5mm}

The subsequent subsections delve into the architecture, design principles, and key features of this C++ framework. They focus on the typing system introduced by Q-Pragma, the concept of quantum routine and the directives extending the C++ language.

\newpage


\subsection{Q-Pragma: A quantum library}

Q-Pragma provides a C++ library defining new quantum types as well as the concept of quantum routines. The new quantum types guarantee compliance with the \textit{typing} and \textit{safe uncomputation} requirements, while the quantum routine concept meets the \textit{reversibility} and \textit{controllability} requirements.


\subsubsection{Quantum types}

In order to make Q-Pragma intuitive and well integrated with classical C++, new quantum types have been added to the framework. These quantum types can describe arbitrary data-structures, including user defined quantum structures. Q-Pragma quantum types can be grouped into two categories:

\begin{itemize}
    \item Quantum boolean, named ``qbool'' in Q-Pragma. A quantum boolean can be seen as a unique pointer, pointing to a quantum memory address. This ``qbool'' is a quantum register of size 1 and is the elementary brick of the quantum type design.
    \item Quantum array, an array of quantum bools having a fixed size. Pure quantum types defined in this framework inherit from a quantum array.
\end{itemize}

A pure quantum register has a fixed size. This size is defined by a constant value and not by a classical variable stored in a classical register, thereby making the register a pure quantum object. This follows the philosophy of C++ types: any C++ type has a fixed size. Moreover, pure quantum functions built upon these fixed-size quantum registers can be optimized at compile-time.
Dynamic-sized quantum types are built using a classical C++ container (e.g. vector, list). They are composed of both classical and quantum registers, and as such are called hybrid structures.

\vspace{5mm}

The safe uncomputation concept stems from the reversibility of quantum unitary operations to reset a quantum register to the $\ket{0}$ state. It has been introduced in some quantum frameworks like Quipper \cite{quipper} or Silq \cite{silq}. Q-Pragma provides a safe uncomputation mechanism based on the initialization of a quantum object. Each quantum type handles its own ``initialization'' (i.e. computing the state preparation - done by the constructor of the quantum type) and ``deinitialization'' (i.e. uncomputing the state preparation - done by the destructor of the quantum type, ignored if a measurement occurred on this register). Any quantum operation executed on the quantum register must be uncomputed to get $\ket{0}$ state. Through a \textit{get\_init} method one can cast a classical type value into a state preparation routine. This routine is used to initialize the state and then stored to undo this state preparation at delete time.

Listing \ref{lst:quantum_type:uncomputation} illustrates the safe uncomputation mechanism. The \textit{qreg} quantum register is initialized to state $\ket{1}$ (to do so, an implicit $X$ gate is applied on the qubit). At the end of the scope, the \textit{qreg} register is no longer accessible, so Q-Pragma reset the register to state $\ket{0}$ (by applying a $X$ gate on the qubit).

\begin{code}
\begin{lstlisting}[language=C++, caption={Safe uncomputation mechanism}, label={lst:quantum_type:uncomputation}]
{
    // Initialize a quantum register based on
    // a classical value
    // This initialization creates a state
    // preparation (here applying a X gate)
    |\qbool|qreg = true;
    
    ...

    // Object "qreg" is deleted, the safe
    // uncomputation mechanism applies a
    // X gate to reset this register to |$\color{colorcomment}{\ket{0}}$|
}
\end{lstlisting}
\end{code}

Similarly, the \textit{cast\_measure} method consists in casting a quantum measurement into a classical interpretation of it. This approach forms an explicit connection between classical and quantum types. For instance, measuring a qbool gives a bool while measuring a quantum array leads to a classical array of bools. If one creates a new quantum type, one can arbitrarily choose the way classical and quantum data interact through this new type.

\begin{code}
\begin{lstlisting}[language=C++, caption={Casting a quantum integer into a classical integer - \textit{A quantum integer being an array of qbools, casting the measurement consists in translating an array of bool (i.e. bitstring) into a classical integer}}, label={lst:quantum_type:measurement}]
// State preparation
|\quinteight|quint = 12UL;

...

// Measurement of a quantum integer
|\cuint{8}|cuint = measure_and_reset(quint);
\end{lstlisting}
\end{code}

Q-Pragma provides three built-in quantum types: the \textit{qbool} (quantum bool or qubit), the \textit{quint\_t} (quantum unsigned integer) and the \textit{qint\_t} (quantum signed integer). The \textit{quint\_t} and the \textit{qint\_t} are inspired from the C++ classical types \textit{uint\_t} and \textit{int\_t} respectively. These two types belong to the \textit{quantum array} category, meaning that \textit{quint\_t} and \textit{qint\_t} are fixed-sized arrays of qbools, their size being given by a template parameter. Note that any quantum type can be created following the \textit{quint\_t} conception.

A quantum integer can be initialized from a classical integer, and measuring a quantum integer returns a classical integer. Listing \ref{lst:quantum_type:measurement} illustrate this behavior, the quantum register \textit{quint} will be initialized with the quantum state $\ket{00001100}$, which is the binary encoding for $12$; at the end of the scope, the quantum register \textit{quint} is measured in the $Z$-basis, and the outcome bitstring is casted into a classical integer. To implement this behavior, quantum type \textit{quint\_t} overloads the \textit{get\_init} and \textit{cast\_measure} methods, enabling this type to be instantiated from an unsigned integer or a positive signed integer.

\begin{code}
\begin{lstlisting}[language=C++, caption={Usage of a 8 qubits quantum integer}, label={lst:quantum_int}]
/* QUINT_T */
|\quinteight|quint_a;
|\quinteight|quint_b = 42;

// Apply operations
quint_a ^= (quint_b + 7);

// Measure a register
|\cuint{8}|cuint_a = measure_and_reset(quint_a);
\end{lstlisting}
\end{code}

A \textit{quint\_t} behaves as a classical integer, enabling bitwise operations and some arithmetic operations, as shown in Listing \ref{lst:quantum_int}. This can be done by overriding C++ operators. For instance, overriding the ``+'' operator enables the addition between \textit{quint\_t} and other quantum or classical integers. The same mechanism applies for \textit{qbool} and \textit{qint\_t}.

\vspace{5mm}

Any built-in Q-Pragma quantum type behaves as its equivalent classical type. A quantum type can be initialized using its equivalent classical type and is manipulated using the same set of operations, as long as the operations remain reversible, as shown in Listing \ref{lst:quantum_int} or Listing \ref{lst:quantum_formula}. Measuring a quantum type returns its classical equivalent. Moreover, any quantum type can be initialized using a C++ expression.

\begin{code}
\begin{lstlisting}[language=C++, caption={Initialization of a quantum type using a C++ expression}, label={lst:quantum_formula}]
// Initializing two qbools from classical bools
|\qbool|a = true, b = false;
// Initializing a qbool from a C++ expression 
|\qbool|c = a |\textbar| b;
\end{lstlisting}
\end{code}

C++ operators can be used to define complex quantum routines using only few lines of code. In addition, this section shows that Q-Pragma adheres to several requirements described in the previous section, namely the \textit{typing} and \textit{safe uncomputation} requirements.


\subsubsection{Quantum routines}

A quantum routine is a pure quantum function acting on a fixed-size quantum register. It is by definition controllable and reversible. A quantum routine is either:

\begin{itemize}
    \item A native gate, also called ``basic gate'' in Q-Pragma. These basic gates are the usual gates used in quantum computation, including Pauli gates X, Y and Z, Toffoli gates, CNOT gates, and others.
    \item A sequence of other quantum routines, each acting on a subset of qubits.
\end{itemize}

A routine being a function, it can be built using the same syntax as any C++ function. A directive \pragma{quantum routine} must be used to declare this function as a quantum routine. The behavior of this pragma is explained in the next section.

\vspace{5mm}

A quantum routine can be executed using the \emph{operator()} method, to be called using the same syntax as any C++ function. Nevertheless, a quantum routine features extra methods to change the behavior of this function, as shown in Listing \ref{lst:base_routine}:

\begin{itemize}
    \item Method \emph{dag()} is used to call the inverse version of the routine.
    \item Method \emph{ctrl()} is used to call the control version of the routine. The size of the register passed in argument defines the number of controls.
    \item Method \emph{ctrl\_dag()} is used to call the control version of the inverse. The size of the register passed in argument defines the number of controls.
\end{itemize}

\begin{code}
\begin{lstlisting}[language=C++, caption={Usage of a quantum routine}, label={lst:base_routine}]
|\qbool |q0, q1, qc;

// Apply the routine
my_routine(q0, q1);

// Apply the reverse routine
my_routine.dag(q0, q1);

// Apply the controlled routine
// The routine is controlled by a single qubit
my_routine.ctrl(qc, q0, q1);

// Apply the reverse and controlled routine
my_routine.ctrl_dag(qc, q0, q1);
\end{lstlisting}
\end{code}

The quantum routine ensures that Q-Pragma supports the \textit{reversibility} and \textit{controllability} requirements.

\newpage


\subsection{Q-Pragma: extending C++}

C++ pragma directives are powerful tools used to specify several compilation features, while maintaining compatibility with the C++ language. Q-Pragma relies on these directives to add quantum programming capabilities to C++. This is achieved using five new pragma directives, allowing hybrid quantum-classical code implementation while remaining consistent with C++, and comprehensible by C++ developers.

\vspace{2mm}

\noindent These pragmas are used to:

\begin{itemize}
    \item manage code and memory locality, i.e. \textit{quantum scope} and \textit{quantum move} directives.
    \item control quantum instructions, i.e. \textit{quantum ctrl} directive.
    \item define quantum routines, i.e. \textit{quantum routine} directive.
    \item manage the uncomputation, i.e. \textit{quantum compute} directive.
\end{itemize}


\subsubsection{Managing code and memory locality}
\label{sec:locality}

The \textit{pragma quantum scope} is the first pragma directive of our framework. It indicates that the following scope has to be run on the quantum device. The QPU has classical capabilities through the controller (see Figure \ref{fig:node-design}). Q-Pragma is based on the assumption that this controller is programmable. Therefore, a quantum scope can contain pure quantum operations, classical instructions, and non-reversible quantum operations like measurements. The \textit{pragma quantum scope} compiles the source code to be offloaded on the QPU, reducing the amount of data transferred, and thus increasing performances. Any function defined on the Host can be called from the quantum device, and a function defined on the quantum device can be called from the Host (quantum gates can then be executed from the Host). This involves a PCI Express communication between QPU and Host, and is therefore slower than executing a local function.

\begin{code}
\begin{lstlisting}[language=C++, caption={Example of a \textit{pragma quantum scope}}, label={lst:pragma_scope}]
|\pragma{quantum scope}|
{
    std::array<int, 4UL> counters;
    |\qbool|q0, q1;
    
    for (int i = 0 ; i < 100 ; ++i) {
        // Generate a Bell pair
        H(q0);
        CNOT(q0, q1);
        
        // Measure
        bool meas0 = measure_and_reset(q0);
        bool meas1 = measure_and_reset(q1);
        
        // Store measurement in the map counter
        ++counters[meas1 << 1 + meas0];
    }
}
\end{lstlisting}
\end{code}

Listing \ref{lst:pragma_scope} illustrates a \textit{pragma quantum scope} directive: the sample of code inside the scope is executed on the QPU, not on the Host. This pragma directive is not mandatory and does not change the program output, but serves to improve performances: the Host communicates only once with the quantum device (to start the scope) instead of a hundred times (to allocate memory, to apply quantum gates or to measure a qubit).

\vspace{5mm}

Any piece of code written inside a \textit{pragma quantum scope} directive is executed on the quantum device, otherwise the code is executed on the Host. If a variable is defined in a \textit{pragma quantum scope} directive, the memory is allocated on the quantum device, otherwise the memory is allocated on the Host. A \textit{qbool} can be allocated and manipulated from the Host, as this structure is a classical object pointing to a quantum memory address (a \textit{qbool} can be seen as a unique pointer). Host variables can be accessed in a \textit{pragma quantum scope}, using the memory sharing enabled by the PCI Express protocol. Nevertheless, variable access between Host and QPU is inherently slow. To compensate for this, Q-Pragma handles the memory through data management directives. Some variables can be temporarily moved from the Host to the QPU in a quantum scope, using the \textit{with} keyword. At the beginning of the scope, the listed variables are moved to the device, and moved back to the Host at the end of the scope. Listing \ref{lst:pragma_scope_with} shows how to temporarily move the \textit{my\_int} variable on the QPU.

\begin{code}
\begin{lstlisting}[language=C++, caption={Example of a \textit{pragma quantum scope with}}, label={lst:pragma_scope_with}]
|\qinteight|my_int;

|\pragma{quantum scope with (my\_int)}|
{
    my_int += 13;
}
\end{lstlisting}
\end{code}

Moreover, the \textit{pragma quantum move} directive handles manual data transfer from Host to device through the \textit{toDevice} specifier, and from device to Host through the \textit{toHost} one. In this case, the developer has total control over the data flow between Host and device as shown in Listing \ref{lst:pragma_move}.

\begin{code}
\begin{lstlisting}[language=C++, caption={Example of a \textit{pragma quantum move} directive}, label={lst:pragma_move}]
std::array<qbool, 8UL> qreg;
bool condition = ...;

|\pragma{quantum scope}|
{
    if (condition) {
        |\pragma{quantum move toDevice(qreg)}|
        for (const auto & qubit : qreg) {
            H(qubit);
        }
        |\pragma{quantum move toHost(qreg)}|
    } else {
        // DO SOMETHING with no memory movement
    }
}
\end{lstlisting}
\end{code}

These two pragma directives provide fine control over both code and memory locality. One can precisely select on which device a function is executed, or a variable stored. The syntaxes of these pragma directives are specified in Figure \ref{fig:pragma_scope_grammar}.

\begin{figure}[H]
    \grammar{
        \nonterm{pragma\_scope}   & ::= & \term{\#pragma quantum scope} {\nonterm{scope\_options}}? \\
        \nonterm{scope\_options}  & ::= & \term{with} \nonterm{var\_list} \\
        \\
        \nonterm{pragma\_move}    & ::= & \term{\#pragma quantum move} (\nonterm{move\_dir} \nonterm{var\_list})+\\
        \nonterm{move\_dir} & ::= & \term{toDevice} $\vert$ \term{toHost} \\
        \\
        \nonterm{var\_list}       & ::= & \term{(} \nontermbutterm{var\_name} (\term{,} \nontermbutterm{var\_name} )* \term{)}
    }
    \caption{Grammar of \textit{pragma quantum scope} and \textit{pragma quantum move} using BNF with regex}
    \label{fig:pragma_scope_grammar}
\end{figure}


\subsubsection{Controlling quantum instructions}

The \textit{pragma quantum ctrl} directive is used to control quantum instructions, using a \textit{qbool} or a C++ expression. If a C++ expression is used, a temporary \textit{qbool} is created and initialized using the C++ expression. Listing \ref{lst:pragma_ctrl} highlights this behavior: a temporary \textit{qbool} is instantiated (with the value $my\_int == 42$), and this \textit{qbool} is used to control the scope; thanks to the \textit{safe uncomputation} mechanism, this temporary \textit{qbool} is reset to state $\ket{0}$ at the end of the scope.

\begin{code}
\begin{lstlisting}[language=C++, caption={Example of a \textit{pragma quantum ctrl} controlled by a C++ expression}, label={lst:pragma_ctrl}]
|\qinteight|my_int;
// Some code acting on my_int
// ...

|\qbool|q0, q1;
|\pragma{quantum ctrl (my\_int == 42)}|
{
    H(q0);
    CNOT(q0,q1);
    RZ(M_PI / 4.)(q1);
}
\end{lstlisting}
\end{code}

Only pure quantum operations are impacted by a control. Control have no impact on classical operations, and quantum measurements are forbidden. A \textit{pragma quantum ctrl} is similar to a quantum \textit{if} statement in the sense that the controlled quantum instructions affect the quantum state only if the control qubit evaluates to \textit{true}. Nevertheless, unlike the C++ \textit{if} statement, instructions inside a \textit{pragma quantum ctrl} are always executed in their controlled version. Using a different syntax between the C++ \textit{if} statement and the \textit{pragma quantum ctrl} avoids confusion. The syntax of this pragma is defined by Figure \ref{fig:pragma_ctrl_gram}.

\begin{figure}[h]
    \grammar{
        \nonterm{pragma\_ctrl}  & ::= & \term{\#pragma quantum ctrl} \nonterm{ctrl\_options} \\
        \nonterm{ctrl\_options} & ::= & \term{(} \nontermbutterm{var\_name} $\vert$ \nontermbutterm{quantum\_condition} \term{)} \\
    }
    \caption{Grammar of \textit{pragma quantum ctrl} using BNF with regex}
    \label{fig:pragma_ctrl_gram}
\end{figure}

This pragma meets the \textit{controllability} requirement of a quantum-classical framework compatible with HPC systems.


\subsubsection{Defining quantum routines}

The \textit{pragma quantum routine} directive is used to create quantum routines using the same syntax as a C++ function. A quantum routine is composed of purely quantum operations, and is reversible. Therefore:

\begin{itemize}
    \item a routine takes only purely quantum objects as arguments.
    \item measurement-based operations are not allowed.
    \item a routine returns nothing (i.e. is void type), since no classical data is either passed as input or created (no measurement).
\end{itemize}

To meet the \textit{reversibility} and \textit{controllability} requirements, a routine is callable, reversible and controllable, as shown in Listing \ref{lst:pragma_routine}. Since a quantum routine is a pure quantum function, the routine is compiled for the QPU: executing a quantum routine from the Host requires a single communication request between the Host and the QPU, regardless of the quantum routine size.

\begin{code}
\begin{lstlisting}[language=C++, caption={Example of the creation and calls of a quantum routine using the \textit{pragma quantum routine} directive}, label={lst:pragma_routine}]
|\pragma{quantum routine}|
void my_routine(const |\qbool|& q0, const |\qbool|& q1)
{
    H(q0);
    Z.ctrl(q0, q1);
}

int main() {
    |\qbool|q0, q1, qc;
    // Apply the routine
    my_routine(q0, q1);
    // Apply the reverse routine
    my_routine.dag(q0, q1);
    // Apply the controlled routine
    my_routine.ctrl(qc, q0, q1);
    // Apply the reverse and controlled routine
    my_routine.ctrl_dag(qc, q0, q1);
}
\end{lstlisting}
\end{code}

In Listing \ref{lst:pragma_routine}, the \textit{my\_routine} routine is compiled for the QPU, i.e. the routine object only exists on the QPU. The \textit{main} function allocates three qubits and performs four calls to \textit{my\_routine}. Since the sample of code is executed on the Host (due to the lack of \textit{pragma quantum scope} directive), each allocation and each routine call sends a request to the QPU. A response is sent back by the QPU after each request, to notify the Host that the request has been executed on the QPU (so the Host can execute the next instruction). Then, executing Listing \ref{lst:pragma_routine} requires 7 requests to the QPU.

\vspace{5mm}

As any C++ function, quantum routines can be templated. For instance, one can create a templated Quantum Fourier Transform (QFT), as shown in Listing \ref{lst::qft}. The template parameter \textit{SIZE} is reused to define the size of the argument \textit{qreg}, but also inside the routine as a stop condition for the \textit{for} loops.

\begin{code}
\begin{lstlisting}[language=C++, caption={QFT on an array, using Q-Pragma}, label={lst::qft}]
|\pragma{quantum routine}|
template <uint64_t SIZE>
void qft(const std::array<qbool, SIZE> & qreg) {
    for (uint64_t idx = 0ul ; idx < SIZE ; ++idx) {
        H(qreg[idx]);
        
        for (uint64_t ctr = idx + 1 ; ctr < SIZE ;
             ++ctr) {
            // Compute angle
            double angle =
                M_PI / (1 << (ctr - idx));

            // Apply gate
            PH(angle).ctrl(qreg[ctr], qreg[idx]);
        }
    }
}

int main() {
    |\qinteight|qreg;
    qft<8UL>(qreg);
}
\end{lstlisting}
\end{code}

Moreover, quantum routines can be parameterized. According to our routine definition, the $R_X$ gate is not a routine as it depends on a classical angle (a routine takes only pure quantum objects as argument). On the other hand, $R_X(\theta)$ is a quantum routine since its angle is known, and the gate itself does not depend on any classical input. In Q-Pragma, $R_X$ is a parameterized routine, i.e. an object which, binded with classical parameters, becomes a routine. Variational algorithms can be implemented with parameterized routine in Q-Pragma.

A parameterized routine can be built using the \textit{pragma quantum routine}, by listing the binded variables in the directive. Listing \ref{lst::routine_param} provides an example of a parameterized routine.

\begin{code}
\begin{lstlisting}[language=C++, caption={Parameterized routine using the \textit{pragma quantum routine}}, label={lst::routine_param}]
|\pragma{quantum routine (\double angle0, \double angle1)}|
void param_routine(const |\qbool|& q0,
                   const |\qbool|& q1) {
    RX(angle0).ctrl(q0, q1);
    RZ(angle1).ctrl(q0, q1);
}

int main() {
    |\qbool|q0, q1;
    param_routine(M_PI/3.0, M_PI/6.0)(q0, q1);
}
\end{lstlisting}
\end{code}

The requirements introduced with the concept of typed arguments in C++ functions do not necessarily apply to quantum routines generated with a \textit{pragma quantum routine}. In fact, as long as the quantum arguments have exactly the same size as the one given in the function declaration, no error will be raised. This brings flexibility, as shown in Listing \ref{lst::routine_flex}.

\begin{code}
\begin{lstlisting}[language=C++, caption={Example demonstrating the flexibility of quantum routine call arguments}, label={lst::routine_flex}]
|\pragma{quantum routine}|
void bell_pair(const |\qbool|& q0, const |\qbool|& q1)
{
    H(q0)
    CNOT(q0, q1);
}

int main() {
    |\qbool|q0, q1;
    bell_pair(q0, q1);  // Correct
    
    std::array<qbool, 2> qreg;
    bell_pair(qreg);  // Still correct
    
    |\qint{2}|qint;
    bell_pair(qint);  // Still correct
}
\end{lstlisting}
\end{code}

However, strict typing can be enforced over the quantum routines using the \textit{typed} flag. This flag ensures that the quantum types given in the quantum routine declaration are exactly the same as the one given in the routine call, as shown in Listing \ref{lst::routine_typed}.

\begin{code}
\begin{lstlisting}[language=C++, caption={Example of a typed quantum routine}, label={lst::routine_typed}]
|\pragma{quantum routine typed}|
void bell_pair(const |\qbool|& q0, const |\qbool|& q1)
{
    H(q0)
    CNOT(q0, q1);
}

int main() {
    |\qbool|q0, q1;
    bell_pair(q0, q1);  // Correct
    
    std::array<qbool, 2> qreg;
    bell_pair(qreg);  // ERROR - does not compile
    
    |\qint{2}|qint;
    bell_pair(qint);  // ERROR - does not compile
}
\end{lstlisting}
\end{code}

Additionally, a pure quantum type has a static size in Q-Pragma. Therefore, quantum routines cannot be created using hybrid types such as lists of qubits. To overcome this limitation, dynamic quantum routines have been introduced to create routines acting on a dynamic-sized quantum registers. A dynamic routine can be created using the ``dynamic'' flag, as shown in Listing \ref{lst::routine_dynamic}. Dynamic routines are always typed.

\begin{code}
\begin{lstlisting}[language=C++, caption={Example of a dynamic quantum routine}, label={lst::routine_dynamic}]
|\pragma{quantum routine dynamic (\double angle)}|
void rx_wall(const std::vector<qbool> & vector) {
    for (auto & qubit: vector) {
        (RX(angle))(qubit);
    }
}

int main() {
    std::vector<qbool> qvect(12);
    rx_wall(0.6)(qvect);
}
\end{lstlisting}
\end{code}

Since dynamic and parameterized quantum routines rely on variables defined at runtime, they can only be created at runtime. Fixed-sized quantum routines are created at compile-time, and are then optimized by the compiler. The syntax of \textit{pragma quantum routine} is defined in Figure \ref{fig:pragma_routine_grammar}.

\begin{figure}[H]
    \grammar{
        \nonterm{prag\_routine} & ::= & \term{\#pragma quantum routine} {\nonterm{flag}}? {\nonterm{bound\_vars}}? \\
        \nonterm{flag}  & ::= & \term{dynamic} $\vert$ \term{typed} \\
        \nonterm{bound\_vars}   & ::= & \term{(} \nonterm{variable} ( \term{,} \nonterm{variable} )* \term{)} \\
        \nonterm{variable}    & ::= & \nontermbutterm{cpp\_type} \nontermbutterm{var\_name} \\
    }
    \caption{Grammar of \textit{pragma quantum routine} using BNF with regex}
    \label{fig:pragma_routine_grammar}
\end{figure}


\subsubsection{Managing the uncomputation}

The \textit{safe uncomputation} requirement is a key element of a quantum-classical framework. Q-Pragma provides a \textit{pragma quantum compute} directive in which pure quantum gates are executed. This directive enables the automatic uncomputation of these pure quantum operations as soon as the end of the current scope is reached. For instance, implementing a $R_{ZZ}$ gate relies on an ancilla qubit which must be reset to its $\ket{0}$ state, this can be implemented using \textit{pragma quantum compute} as shown in Listing \ref{lst::compute}.

\begin{code}
\begin{lstlisting}[language=C++, caption={Example of a parameterized routine using the \textit{pragma quantum routine} directive}, label={lst::compute}]
|\pragma{quantum routine (\double angle)}|
void RZZ(const |\qbool|& qb1, const |\qbool|& qb2) {
    // Define ancilla
    |\qbool|ancilla;

    {
        |\pragma{quantum compute}|
        {
            CCNOT(qb1, qb2, ancilla);
        }

        // Apply RZ gate
        (RZ(angle))(ancilla);

        // Automatic uncomputation of CCNOT
        // "ancilla" is then reset to |$\color{colorcomment}{\ket{0}}$|
    }
}
\end{lstlisting}
\end{code}

This directive has been designed to meet the \textit{safe uncomputation} requirement, but can also be used to implement a routine matching the pattern $U \cdot A \cdot U^\dagger$. This allows additional optimization techniques. For instance, the computation and uncomputation parts are never controlled, even if they are in a \textit{pragma quantum ctrl} context. The syntax of this directive is shown in Figure \ref{fig:pragma_compute_gram}.

\begin{figure}[h]
    \grammar{
        \nonterm{pragma\_compute} & ::= & \term{\#pragma quantum compute}\\
    }
    \caption{Grammar of \textit{pragma quantum compute} using BNF with regex}
    \label{fig:pragma_compute_gram}
\end{figure}

\newpage


\section{Evaluation of Q-Pragma}

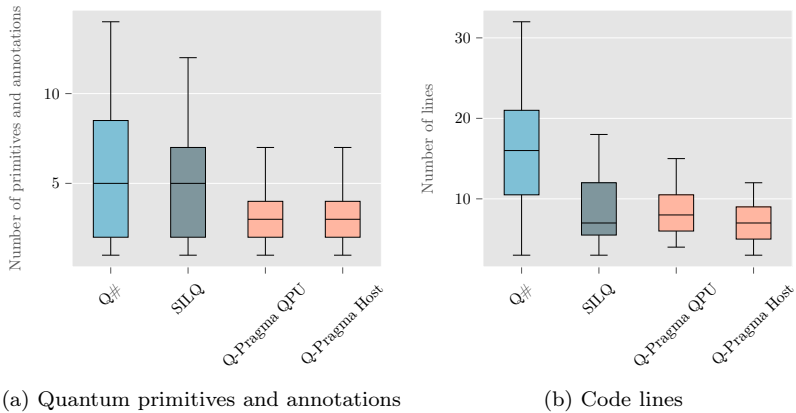
\begin{figure}[!h]
    \centering
    \subfloat[Quantum primitives and annotations]{
        \scalebox{0.6}{
\begin{tikzpicture}

\definecolor{dimgray85}{RGB}{85,85,85}
\definecolor{gainsboro229}{RGB}{229,229,229}
\definecolor{lightpink255182161}{RGB}{255,182,161}
\definecolor{lightslategray127150157}{RGB}{127,150,157}
\definecolor{skyblue127192214}{RGB}{127,192,214}

\begin{axis}[
axis background/.style={fill=gainsboro229},
axis line style={white},
tick align=outside,
tick pos=left,
x grid style={white},
xmin=0.5, xmax=4.5,
xtick style={color=dimgray85},
xtick={1,2,3,4},
xticklabel style={rotate=45.0},
xticklabels={Q\#,SILQ,Q-Pragma QPU,Q-Pragma Host},
y grid style={white},
ylabel=\textcolor{dimgray85}{Number of primitives and annotations},
ymajorgrids,
ymin=0.35, ymax=14.65,
ytick style={color=dimgray85}
]
\path [draw=black, fill=skyblue127192214]
(axis cs:0.775,2)
--(axis cs:1.225,2)
--(axis cs:1.225,8.5)
--(axis cs:0.775,8.5)
--(axis cs:0.775,2)
--cycle;
\addplot [black]
table {%
1 2
1 1
};
\addplot [black]
table {%
1 8.5
1 14
};
\addplot [black]
table {%
0.8875 1
1.1125 1
};
\addplot [black]
table {%
0.8875 14
1.1125 14
};
\path [draw=black, fill=lightslategray127150157]
(axis cs:1.775,2)
--(axis cs:2.225,2)
--(axis cs:2.225,7)
--(axis cs:1.775,7)
--(axis cs:1.775,2)
--cycle;
\addplot [black]
table {%
2 2
2 1
};
\addplot [black]
table {%
2 7
2 12
};
\addplot [black]
table {%
1.8875 1
2.1125 1
};
\addplot [black]
table {%
1.8875 12
2.1125 12
};
\path [draw=black, fill=lightpink255182161]
(axis cs:2.775,2)
--(axis cs:3.225,2)
--(axis cs:3.225,4)
--(axis cs:2.775,4)
--(axis cs:2.775,2)
--cycle;
\addplot [black]
table {%
3 2
3 1
};
\addplot [black]
table {%
3 4
3 7
};
\addplot [black]
table {%
2.8875 1
3.1125 1
};
\addplot [black]
table {%
2.8875 7
3.1125 7
};
\path [draw=black, fill=lightpink255182161]
(axis cs:3.775,2)
--(axis cs:4.225,2)
--(axis cs:4.225,4)
--(axis cs:3.775,4)
--(axis cs:3.775,2)
--cycle;
\addplot [black]
table {%
4 2
4 1
};
\addplot [black]
table {%
4 4
4 7
};
\addplot [black]
table {%
3.8875 1
4.1125 1
};
\addplot [black]
table {%
3.8875 7
4.1125 7
};
\addplot [black]
table {%
0.775 5
1.225 5
};
\addplot [black]
table {%
1.775 5
2.225 5
};
\addplot [black]
table {%
2.775 3
3.225 3
};
\addplot [black]
table {%
3.775 3
4.225 3
};
\end{axis}

\end{tikzpicture}}
        \label{fig:bench:primitives}
    }
    \subfloat[Code lines]{
        \scalebox{0.6}{
\begin{tikzpicture}

\definecolor{dimgray85}{RGB}{85,85,85}
\definecolor{gainsboro229}{RGB}{229,229,229}
\definecolor{lightpink255182161}{RGB}{255,182,161}
\definecolor{lightslategray127150157}{RGB}{127,150,157}
\definecolor{skyblue127192214}{RGB}{127,192,214}

\begin{axis}[
axis background/.style={fill=gainsboro229},
axis line style={white},
tick align=outside,
tick pos=left,
x grid style={white},
xmin=0.5, xmax=4.5,
xtick style={color=dimgray85},
xtick={1,2,3,4},
xticklabel style={rotate=45.0},
xticklabels={Q\#,SILQ,Q-Pragma QPU,Q-Pragma Host},
y grid style={white},
ylabel=\textcolor{dimgray85}{Number of lines},
ymajorgrids,
ymin=1.55, ymax=33.45,
ytick style={color=dimgray85}
]
\path [draw=black, fill=skyblue127192214]
(axis cs:0.775,10.5)
--(axis cs:1.225,10.5)
--(axis cs:1.225,21)
--(axis cs:0.775,21)
--(axis cs:0.775,10.5)
--cycle;
\addplot [black]
table {%
1 10.5
1 3
};
\addplot [black]
table {%
1 21
1 32
};
\addplot [black]
table {%
0.8875 3
1.1125 3
};
\addplot [black]
table {%
0.8875 32
1.1125 32
};
\path [draw=black, fill=lightslategray127150157]
(axis cs:1.775,5.5)
--(axis cs:2.225,5.5)
--(axis cs:2.225,12)
--(axis cs:1.775,12)
--(axis cs:1.775,5.5)
--cycle;
\addplot [black]
table {%
2 5.5
2 3
};
\addplot [black]
table {%
2 12
2 18
};
\addplot [black]
table {%
1.8875 3
2.1125 3
};
\addplot [black]
table {%
1.8875 18
2.1125 18
};
\path [draw=black, fill=lightpink255182161]
(axis cs:2.775,6)
--(axis cs:3.225,6)
--(axis cs:3.225,10.5)
--(axis cs:2.775,10.5)
--(axis cs:2.775,6)
--cycle;
\addplot [black]
table {%
3 6
3 4
};
\addplot [black]
table {%
3 10.5
3 15
};
\addplot [black]
table {%
2.8875 4
3.1125 4
};
\addplot [black]
table {%
2.8875 15
3.1125 15
};
\path [draw=black, fill=lightpink255182161]
(axis cs:3.775,5)
--(axis cs:4.225,5)
--(axis cs:4.225,9)
--(axis cs:3.775,9)
--(axis cs:3.775,5)
--cycle;
\addplot [black]
table {%
4 5
4 3
};
\addplot [black]
table {%
4 9
4 12
};
\addplot [black]
table {%
3.8875 3
4.1125 3
};
\addplot [black]
table {%
3.8875 12
4.1125 12
};
\addplot [black]
table {%
0.775 16
1.225 16
};
\addplot [black]
table {%
1.775 7
2.225 7
};
\addplot [black]
table {%
2.775 8
3.225 8
};
\addplot [black]
table {%
3.775 7
4.225 7
};
\end{axis}

\end{tikzpicture}}
        \label{fig:bench:lines}
    }
    \caption{Evaluation of Q-Pragma compared to Q\# and SILQ. The comparison between these frameworks is based on the number of quantum primitives and annotations used (Figure \ref{fig:bench:primitives}) and the code length (Figure \ref{fig:bench:lines}) - Q-Pragma allows these algorithms to be implemented on the Host or on the QPU (by using a \textit{pragma quantum scope} directive). Each algorithm of this benchmark has been implemented twice (one implementation executing code on the Host, the other executing code on the QPU). See Appendix \ref{app:bench} for details.}
    \label{fig:bench}
\end{figure}

Q-Pragma has been designed to simplify the development of quantum algorithms. The article presenting the SILQ framework \cite{silq} compares some frameworks to show how suitable they are for implementing various quantum algorithms. This benchmark has been extended to add Q-Pragma figures, as shown in Figure \ref{fig:bench}.

Typing provides a higher-level interface for manipulating quantum registers, simplifying the development of quantum algorithms. Thus, complex quantum algorithms can be written using fewer quantum primitives (see Figure \ref{fig:bench:primitives}), and using less line of code (see Figure \ref{fig:bench:lines}).


\section{Conclusion}

Q-Pragma enables quantum code integration in classical applications, and remains compatible with HPC environments. It has been designed as an Open Standard, to ensure that these applications will be executable on real QPUs. This also guarantees that Q-Pragma can evolve, and support new programming features future hybrid code may require. Q-Pragma attempts to meet all the requirements current and future QPUs will establish, facilitating the development of hybrid quantum-classical applications able to reach quantum advantage.

\vspace{5mm}

Since the relevance of a framework can only be assessed through practice, our aim is to develop post-NISQ quantum algorithms using Q-Pragma. For example, we would like to implement several linear algebra algorithms, such as QSVT \cite{QSVT_improvements, Grand_unification}, since their classical equivalents are of great interest in HPC.

\vspace{1mm}

Furthermore, Q-Pragma has only been connected to quantum emulators, and not yet to real quantum hardware. We would like to work with QPU makers to demonstrate a real use case of Q-Pragma.

\vspace{1mm}

Q-Pragma also introduces new compilation constraints. Current quantum compilers process quantum circuits, acting on a known set of qubits. Q-Pragma defines quantum routines, but the register on which a routine will be applied is not known at compile time. This makes it difficult to consider topological constraints when compiling a Q-Pragma program. We would like to work on compilation within the Q-Pragma framework, whether in a QEC scope or not.

\vspace{5mm}

This Q-Pragma framework has been implemented in C++17, and relies on a Clang plugin. Some common quantum algorithms are implemented in Q-Pragma, code examples are given in the Appendix \ref{app:examples}.


\section{Acknowledgement}

\noindent This work is part of \href{https://www.hqi.fr/}{HQI initiative} and is supported by France 2030 under the French National Research Agency award number ``ANR-22-PNCQ-0002''.

\vspace{2mm}

\noindent The authors acknowledge Maxime OLIVA and Stanley CHEAH for their comments and feedbacks, which greatly improved this article.

\vspace{2mm}

\noindent The authors would also like to show their gratitude to Marc BABOULIN and Cyril ALLOUCHE for sharing their insights.

\vspace{2mm}

\noindent The authors sincerely appreciate the support of the \textit{Eviden Quantum Lab} team. Without our valued team, Q-Pragma framework implementation would not have been possible.


\bibliographystyle{plain}
\bibliography{biblio}

\begin{thebibliography}{10}

\bibitem{scaffold}
Ali Abhari, Arvin Faruque, Mohammad~Javad Dousti, Lukas Svec, Oana Catu, Amlan Chakrabati, Chen-Fu Chiang, Seth Vanderwilt, John Black, Frederic Chong, Margaret Martonosi, Martin Suchara, Ken Brown, Massoud Pedram, and Todd Brun.
\newblock Scaffold: Quantum programming language, 07 2012.

\bibitem{lapack}
E.~Anderson, Z.~Bai, C.~Bischof, S.~Blackford, J.~Demmel, J.~Dongarra, J.~Du~Croz, A.~Greenbaum, S.~Hammarling, A.~McKenney, and D.~Sorensen.
\newblock {\em {LAPACK} Users' Guide}.
\newblock Society for Industrial and Applied Mathematics, Philadelphia, PA, third edition, 1999.

\bibitem{fortran}
John Backus.
\newblock The history of fortran i, ii and iii.
\newblock {\em Annals of the History of Computing}, 1(1):21--37, 1979.

\bibitem{quantum_computer_architecture}
K.~Bertels, A.~Sarkar, T.~Hubregtsen., M.~Serrao, A.~A. Mouedenne, A.~Yadav, A.~Krol, and I.~Ashraf.
\newblock Quantum computer architecture: Towards full-stack quantum accelerators.
\newblock In {\em 2020 Design, Automation {\&} Test in Europe Conference {\&} Exhibition (2020)}. {IEEE}, March 2020.

\bibitem{silq}
Benjamin Bichsel, Maximilian Baader, Timon Gehr, and Martin Vechev.
\newblock Silq: a high-level quantum language with safe uncomputation and intuitive semantics.
\newblock In {\em Proceedings of the 41st {ACM} {SIGPLAN} Conference on Programming Language Design and Implementation}. {ACM}, June 2020.

\bibitem{BLAS}
L~Susan Blackford, Antoine Petitet, Roldan Pozo, Karin Remington, R~Clint Whaley, James Demmel, Jack Dongarra, Iain Duff, Sven Hammarling, Greg Henry, et~al.
\newblock An updated set of basic linear algebra subprograms (blas).
\newblock {\em ACM Transactions on Mathematical Software}, 28(2):135--151, 2002.

\bibitem{grid_based_quantum_chemistry}
Hans Hon~Sang Chan, Richard Meister, Tyson Jones, David~P. Tew, and Simon~C. Benjamin.
\newblock Grid-based methods for chemistry simulations on a quantum computer.
\newblock {\em Science Advances}, 9(9), mar 2023.

\bibitem{cirq}
{Cirq Developers}.
\newblock Cirq, 2023.

\bibitem{openqasm3}
Andrew Cross, Ali Javadi-Abhari, Thomas Alexander, Niel~De Beaudrap, Lev~S. Bishop, Steven Heidel, Colm~A. Ryan, Prasahnt Sivarajah, John Smolin, Jay~M. Gambetta, and Blake~R. Johnson.
\newblock {OpenQASM}~3: A broader and deeper quantum assembly language.
\newblock {\em {ACM} Transactions on Quantum Computing}, 3(3):1--50, September 2022.

\bibitem{openmp}
Leonardo Dagum and Ramesh Menon.
\newblock Openmp: an industry standard api for shared-memory programming.
\newblock {\em Computational Science \& Engineering, IEEE}, 5(1):46--55, 1998.

\bibitem{cudaquantum}
NVIDIA CUDA~Quantum development team.
\newblock Cuda quantum, 2023.

\bibitem{DiVincenzo2000}
David~P. DiVincenzo.
\newblock The physical implementation of quantum computation.
\newblock {\em Fortschritte der Physik}, 48(9–11):771–783, September 2000.

\bibitem{myqlm}
{Eviden Quantum Lab}.
\newblock {myQLM: Quantum Computing Framework}, 2020-2023.

\bibitem{WindowedArith}
Craig Gidney.
\newblock Windowed quantum arithmetic.
\newblock {\em arXiv: Quantum Physics}, May 2019.

\bibitem{Shor_8h}
Craig Gidney and Martin Eker{\aa}.
\newblock How to factor 2048 bit {RSA} integers in 8 hours using 20 million noisy qubits.
\newblock {\em Quantum}, 5:433, April 2021.

\bibitem{QSVT_improvements}
Andr{\'{a}}s Gily{\'{e}}n, Yuan Su, Guang~Hao Low, and Nathan Wiebe.
\newblock Quantum singular value transformation and beyond: exponential improvements for quantum matrix arithmetics.
\newblock In {\em Proceedings of the 51st Annual {ACM} {SIGACT} Symposium on Theory of Computing}. {ACM}, 2019.

\bibitem{quipper}
Alexander~S. Green, Peter~LeFanu Lumsdaine, Neil~J. Ross, Peter Selinger, and Beno{\^{\i} }t Valiron.
\newblock Quipper.
\newblock {\em {ACM} {SIGPLAN} Notices}, 48(6):333--342, jun 2013.

\bibitem{ProjectQ_2}
Thomas H\"{a}ner, Damian~S Steiger, Krysta Svore, and Matthias Troyer.
\newblock A software methodology for compiling quantum programs.
\newblock {\em Quantum Science and Technology}, 3(2):020501, February 2018.

\bibitem{gpgpu}
Qihang Huang, Zhiyi Huang, Paul Werstein, and Martin Purvis.
\newblock Gpu as a general purpose computing resource.
\newblock In {\em 2008 Ninth International Conference on Parallel and Distributed Computing, Applications and Technologies}, pages 151--158, 2008.

\bibitem{quantum_computer_for_hpc}
Travis~S. Humble, Alexander McCaskey, Dmitry~I. Lyakh, Meenambika Gowrishankar, Albert Frisch, and Thomas Monz.
\newblock Quantum computers for high-performance computing.
\newblock {\em IEEE Micro}, 41(5):15--23, 2021.

\bibitem{scaffcc}
Ali JavadiAbhari, Shruti Patil, Daniel Kudrow, Jeff Heckey, Alexey Lvov, Frederic~T. Chong, and Margaret Martonosi.
\newblock {ScaffCC}: Scalable compilation and analysis of quantum programs.
\newblock {\em Parallel Computing}, 45:2--17, jun 2015.

\bibitem{quantum_chemistry}
Ivan Kassal, James~D. Whitfield, Alejandro Perdomo-Ortiz, Man-Hong Yung, and Al{\'{a} }n Aspuru-Guzik.
\newblock Simulating chemistry using quantum computers.
\newblock {\em Annual Review of Physical Chemistry}, 62(1):185--207, may 2011.

\bibitem{C}
Brian~W Kernighan and Dennis~M Ritchie.
\newblock {\em The C programming language}.
\newblock Pearson, 2006.

\bibitem{openmp_gpu}
Seyong Lee, Seung-Jai Min, and Rudolf Eigenmann.
\newblock Openmp to gpgpu: A compiler framework for automatic translation and optimization.
\newblock {\em SIGPLAN Not.}, 44(4):101–110, feb 2009.

\bibitem{Grand_unification}
John~M. Martyn, Zane~M. Rossi, Andrew~K. Tan, and Isaac~L. Chuang.
\newblock Grand unification of quantum algorithms.
\newblock {\em {PRX} Quantum}, 2(4), dec 2021.

\bibitem{qcor}
Alexander Mccaskey, Thien Nguyen, Anthony Santana, Daniel Claudino, Tyler Kharazi, and Hal Finkel.
\newblock Extending c++ for heterogeneous quantum-classical computing.
\newblock {\em ACM Transactions on Quantum Computing}, 2(2):1–36, 2021.

\bibitem{S18}
Mariia Mykhailova and Martin Roetteler.
\newblock Microsoft q\# coding contest - summer 2018 - main contest july 6-9, 2018, 2018.

\bibitem{W19}
Mariia Mykhailova and Martin Roetteler.
\newblock Microsoft q\# coding contest - winter 2019 - main contest march 1-4, 2019, 2018.

\bibitem{nielsen_chuang}
Michael Nielsen and Isaac Chuang.
\newblock Quantum information theory.
\newblock In {\em Quantum Computation and Quantum Information}, pages 528--607. Cambridge University Press, June 2012.

\bibitem{cuda}
NVIDIA, Péter Vingelmann, and Frank~H.P. Fitzek.
\newblock Cuda, release: 10.2.89, 2020.

\bibitem{qiskit}
{Qiskit contributors}.
\newblock Qiskit: An open-source framework for quantum computing, 2023.

\bibitem{Shor_1997}
Peter~W. Shor.
\newblock Polynomial-time algorithms for prime factorization and discrete logarithms on a quantum computer.
\newblock {\em {SIAM} Journal on Computing}, 26(5):1484--1509, oct 1997.

\bibitem{ProjectQ_1}
Damian~S. Steiger, Thomas H\"{a}ner, and Matthias Troyer.
\newblock {ProjectQ}: an open source software framework for quantum computing.
\newblock {\em Quantum}, 2:49, January 2018.

\bibitem{openCL}
John~E. Stone, David Gohara, and Guochun Shi.
\newblock {OpenCL}: A parallel programming standard for heterogeneous computing systems.
\newblock {\em Computing in Science {\&} Engineering}, 12(3):66--73, May 2010.

\bibitem{C++}
Bjarne Stroustrup.
\newblock {\em The C++ programming language}.
\newblock Addison-Wesley, 3rd edition, 1997.

\bibitem{Qsharp}
Krysta Svore, Alan Geller, Matthias Troyer, John Azariah, Christopher Granade, Bettina Heim, Vadym Kliuchnikov, Mariia Mykhailova, Andres Paz, and Martin Roetteler.
\newblock Q{\#}.
\newblock In {\em Proceedings of the Real World Domain Specific Languages Workshop 2018}. {ACM}, feb 2018.

\end{thebibliography}


\newpage
\appendix

\section{Examples} \label{app:examples}


\subsection{Bell pair}

The Bell pair algorithm is a quantum routine composed of two quantum gates. These quantum gates are part of Q-Pragma standard library. We provide in Listing \ref{lst:example:bell_pair} an example of such implementation.

\begin{code}
\begin{lstlisting}[language=C++, caption={Example - Bell pair}, label={lst:example:bell_pair}]
|\pragma{quantum routine}|
void bell_pair(const |\qbool|& qb0,
               const |\qbool|& qb1) {
    H(q0);
    CNOT(q0, q1);
}
\end{lstlisting}
\end{code}


\subsection{Uniform superposition}

A uniform superposition $\left[0\;,\;256\right[$ can be created using wall of Hadamard gates acting on $8$ qubits, as $256 = 2^8$. On the other hand, a uniform superposition $\left[0\,,\,200\right[$ requires an additional post-selection, as shown in Listing \ref{lst:example:uniform}.

\begin{code}
\begin{lstlisting}[language=C++, caption={Example - Uniform superposition $\left[0\;,\;200\right[$}, label={lst:example:uniform}]
|\quinteight|quantum_int;
|\qbool|ancilla;

do {
    // Create a uniform superposition
    reset(quantum_int);
    wall::H<8UL>(quantum_int);

    |\pragma{quantum ctrl(quantum\_int >= 200)}|
    X(ancilla);
} while (measure_and_reset(ancilla));
\end{lstlisting}
\end{code}


\subsection{Shor Algorithm}

Shor algorithm \cite{Shor_1997} relies on advanced primitives like the Modular Exponentiation. In Q-Pragma, this primitive can be easily implemented using C++ operators. This is demonstrated in the Shor algorithm implementation in Listing \ref{lst:example:shor}. This example only focuses on the quantum part of Shor algorithm.

\begin{code}
\begin{lstlisting}[language=C++, caption={Example - Shor algorithm}, label={lst:example:shor}]
// Define the size of the register used to encode
// a quantum (unsigned) integer
#define SIZE ...

using namespace qpragma;

uint64_t find_divisor(uint64_t to_divide) {
    // Step 1: find a random number
    std::random_device rd;
    std::uniform_in_distribution<uint64_t> distrib(
        2UL, to_divide - 1UL
    );
    uint64_t random_number = distrib(rd);

    // If random_number is not coprime with
    // to_divide, then we found a divisor
    if (
        auto gcd =
            std::gcd(random_number, to_divide);
        gcd != 1UL
    )
        return gcd;

    // Step 2: Perform the quantum part of Shor
    uint64_t measurement = 0UL;

    |\pragma{quantum scope with (to\_divide, random\_number)}|
    {
        |\quint{SIZE}|first_register;
        wall::H<SIZE>(first_register);

        |\quint{SIZE}|second_register =
            qpragma::pow(random_number,
                         first_register)
            % to_divide;
        reset(second_register);

        qft<SIZE>(first_register);
        measurement =
            measure_and_reset(first_register);
    }

    // Step 3: perform the classical part of
    // Shor algorithm. This part is purely
    // classical and rely on the continuous
    // fraction algorithm
    ...
}
\end{lstlisting}
\end{code}


\newpage

\section{Features implemented in existing frameworks} \label{app:features}

This section explains the reasoning behind Table \ref{tab:frameworks} and is composed of two subsections. The first subsection  focuses on quantum frameworks, and the second subsection focuses on well-established hybrid CPU-GPU frameworks.

\subsection{Quantum frameworks}

\paragraph{Locality}
\textit{A quantum-HPC framework provides tools to control code and memory locality.} \\

\begin{explaination}
    \noindent Most of today’s quantum frameworks do not provide control over QPU-Host interactions, meaning that the code written in these frameworks can only be executed on a quantum device. Therefore, a user cannot choose whether a function should be executed on the Host or on the QPU. Then, most of today's quantum frameworks do not implement the \textit{locality} feature. \\

    \noindent Nevertheless, CUDA Quantum \cite{cudaquantum} and QCOR \cite{qcor} introduce the concept of quantum kernels, to specify that a function should be executed on a quantum device (otherwise, the function is executed on the Host). Nevertheless, the \textit{locality} feature is not entirely supported as these frameworks do not provide tools to control the memory locality.
\end{explaination}

\paragraph{Dynamic interaction}
\textit{Classical memory and quantum memory can be allocated on a QPU and manipulated from the Host, at runtime.} \\

\begin{explaination}
    \noindent Most of quantum frameworks rely on the concept of quantum circuit to describe a quantum computation. In these frameworks, a circuit describes the entire quantum computation: each circuit describes an independent computation, and the quantum memory is reset after the circuit has been executed. Moreover, a quantum circuit being static, a classical computation cannot interact with it during the execution. As a result, quantum frameworks relying on a quantum circuit structure prevent any interaction between the classical and quantum parts of the system, so do not implement the \textit{dynamic interaction} feature. \\

    \noindent In most of quantum frameworks, applying a gate is equivalent to adding it into an underlying circuit, but this is not mandatory. For instance, the instruction can be streamed to the QPU and the gate applied right away, allowing a classical program to interact with the QPU. Several methods have been implemented by OpenQASM 3 (see \cite{openqasm3}, Section 2.4), ProjectQ (see \cite{ProjectQ_1}, Section 4.1.2) and Q\# (see examples from Q\# documentation) to support the \textit{dynamic interaction} feature. \\

    \noindent All other frameworks listed in Table \ref{tab:frameworks} rely on a circuit structure to describe a quantum computation, so do not implement the \textit{dynamic interaction} feature.
\end{explaination}

\paragraph{Scalability}
\textit{Hybrid quantum-HPC frameworks should support algorithms with arbitrary big number of qubits, or instructions.} \\

\begin{explaination}
    \noindent The circuit structure, used by most of today's quantum frameworks, list all quantum operations that will occur during a computation. This structure does not scale well as the number of qubits continues to increase. OpenQASM 3 \cite{openqasm3} and ProjectQ \cite{ProjectQ_1, ProjectQ_2} rely on scalable structures (see \textit{dynamic interaction} paragraph) to describe a quantum computation. \\

    \noindent Q\# \cite{Qsharp} is the only remaining framework which does not rely on a quantum circuit. Nevertheless, due to the lack of information concerning the internal structure used by this framework, it is not possible to verify whether this framework support the \textit{scalability} feature or not.
\end{explaination}

\paragraph{Typing}
\textit{Quantum registers are typed to simplify the manipulation of huge structures.} \\

\begin{explaination}
    \noindent Typing registers add a new level of abstraction to simplify the manipulation of large data structures, simplifying the design of large scale quantum algorithms. NISQ frameworks focusing on short-term algorithms, they often do not implement this feature. As a consequence, only SILQ \cite{silq} and myQLM \cite{myqlm} implement the \textit{typing} feature.
\end{explaination}

\paragraph{Reversibility}
\textit{Pure quantum operations are reversible.} \\

\begin{explaination}
    \noindent The \textit{reversibility} feature is key to simplify the development of quantum algorithms. This feature is already well integrated in existing quantum frameworks. Scaffold \cite{scaffcc} is the only framework listed in Table \ref{tab:frameworks} which does not fulfill this requirement (reversing a routine in Scaffold implies to rewrite the routine by hand, as shown in their tutorials).
\end{explaination}

\paragraph{Controllability}
\textit{Pure quantum operations are controllable.} \\

\begin{explaination}   
    \noindent To simplify the development of quantum-HPC algorithms, a quantum routine should be controllable by one or several qubits. All the programming frameworks listed in Table \ref{tab:frameworks} implement the \textit{controllability} feature, by providing either a control function / method, or by relying on a \textit{if} statement. 
\end{explaination}

\paragraph{Safe uncomputation}
\textit{Quantum register can be reset to $\ket{0}$ state without measurement.} \\

\begin{explaination}    
    \noindent The safe uncomputation concept has been introduced by Quipper \cite{quipper} and embraced by SILQ \cite{silq}. Additional frameworks like CUDA Quantum \cite{cudaquantum}, myQLM \cite{myqlm}, ProjectQ \cite{ProjectQ_1}, or Q\# do not full-fill this requirement but provides some uncomputation features that enable to uncompute safely a quantum register.
\end{explaination}

\subsection{Classical frameworks}

\paragraph{Locality}
\textit{A hybrid framework provides tools to control code and memory locality.}

\begin{explaination}

    \paragraph{CUDA}
    Any function defined in CUDA \cite{cuda} is, by default, executed on the Host. Some function attribute like \textit{\_\_Host\_\_} or \textit{\_\_device\_\_} can be used to specify if a function should be executed on the Host or on the GPU. A Host function is only callable from the Host, and a device function is only callable from the device. The \textit{\_\_global\_\_} function attribute is used to define a kernel, which is a function executed on the device but callable from the Host. By using these attributes, one can control the code locality. \\

    \noindent CUDA is based on C++. Then, any C++ function allocating memory on the Host is available in CUDA. Moreover, CUDA provides some functions, like \textit{cudaMalloc}, to allocate memory on the device. By using C++ and CUDA functions, one can control the memory locality.

    \paragraph{OpenCL}
    An OpenCL-based program is developed using a classical programming language (like C) and OpenCL \cite{openCL}. Any function defined in the classical language is executed on the Host, and any function written in OpenCL is executed on the GPU. OpenCL defines the \textit{kernel} function attribute to define a kernel, which is a function executed on the device but callable from the classical programming language. By using these two programming languages, one can control the code locality. \\

    \noindent OpenCL provides also a library to allocate memory on the GPU, using a classical programming language, to control the memory locality.

    \paragraph{OpenMP}
    OpenMP \cite{openmp, openmp_gpu} extends C++. Any C++ function is executed on the Host. OpenMP relies on the \textit{omp target} pragma directive to define a kernel, which is a piece of code executed on the device. By using this pragma directive, one can control the code locality. \\

    \noindent OpenMP does not provide tools to directly allocate memory on the GPU. Nevertheless, the \textit{omp target} pragma directive can move data from the Host to the device (and vice-versa), to control the memory locality.
\end{explaination}

\paragraph{Dynamic interaction}
\textit{Classical memory and GPU memory can be allocated on a GPU and manipulated from the Host, at runtime.} \\

\begin{explaination}
    \noindent As shown above, CUDA \cite{cuda}, OpenCL \cite{openCL}, and OpenMP \cite{openmp, openmp_gpu} provide tools to allocate memory at runtime. This memory can be updated at runtime like any Host variable (the GPU being connected using a PCIe link, the GPU memory is mapped and accessible from the Host).

    \noindent Moreover, kernels can be started at runtime, fulfilling the \textit{dynamic interaction} feature.
\end{explaination}

\paragraph{Scalability}
\textit{Hybrid CPU-GPU frameworks should support algorithms with arbitrary memory size, or instructions.} \\

\begin{explaination}
    \noindent CUDA \cite{cuda}, OpenCL \cite{openCL}, and OpenMP \cite{openmp, openmp_gpu} are used in production to solve complex problems requiring the manipulation of large amounts of data. The \textit{scalability} feature is already proven by practice.
\end{explaination}

\paragraph{Typing}
\textit{GPU registers are typed to simplify the manipulation of huge structures.} \\

\begin{explaination}
    \noindent CUDA \cite{cuda}, and OpenMP \cite{openmp, openmp_gpu} are based on the C++ language. Any C++ class or type can be used on the GPU using these frameworks. \\

    \noindent OpenCL \cite{openCL} is based on C language. Any C type is already defined in OpenCL.
\end{explaination}


\newpage

\section{Evaluation of Q-Pragma} \label{app:bench}
This paper provides a first evaluation of Q-Pragma compared to Q\# \cite{Qsharp} and SILQ \cite{silq}. This comparison shows that implementing algorithms using Q-Pragma requires less lines of code, in average. Table \ref{tab:bench} summarizes the code length, and the number of quantum primitives / annotations needed to solve various problems defined for the Microsoft Q\# coding contests (Summer 2018 \cite{S18} and Winter 2019 \cite{W19}). This evaluation has already be done in \cite{silq} for both Q\# and SILQ.

\begin{table}[h]
    \centering
    \subfloat[Summer 2018 comparison]{
        \scalebox{0.6}{\begin{tabular}{|c|l|c|c|c|c|c|c|c|c|c|c|c|c|c|c|c|}
    \cline{3-17}
    \multicolumn{2}{c|}{} & \multicolumn{15}{c|}{Summer 2018}  \\
    \cline{3-17}
    \multicolumn{2}{c|}{} & A1 & A2 & A3 & A4 & B1 & B2 & B3 & B4 & C1 & C2 & D1 & D2 & D3 & E1 & E2 \\
    \hline
    \multirow{2}{*}{Q\#} & Lines & 9 & 12 & 32 & 24 & 12 & 16 & 9 & 19 & 11 & 28 & 11 & 15 & 9 & 23 & 21 \\
                         \cline{2-17}
                         & QPA & \best 1 & \best 2 & \best 4 & 6 & \best 1 & \best 1 & 5 & 14 & \best 2 & \best 3 & \best 1 & 4 & 3 & 8 & \best 3 \\
    \hline
    \multirow{2}{*}{SILQ} & lines & 5 & \best 6 & 12 & 12 & \best 3 & 9 & \best 4 & \best 5 & \best 3 & \best 7 & 7 & 7 & 7 & \best 7 & 5 \\
                          \cline{2-17}
                          & QPA & 2 & 5 & 5 & 7 & 2 & 2 & \best 3 & 5 & \best 2 & 5 & 3 & 3 & \best 1 & 9 & 5 \\
    \hline
    \multirow{2}{*}{Q-Pragma} & Lines & \best 3+1 & \best 6+1 & \best 8+1 & \best 9+1 & 4+3 & \best 5+3 & \best 4+3 & 9+3 & 4+3 & 9+3 & \best 5+1 & \best 5+1 & \best 6+0 & 8+3 & 9+4 \\
                              \cline{2-17}
                              & QPA & 2+0 & 3+0 & \best 4+0 & \best 2+1 & \best 1+0 & \best 1+0 & \best 2+1 &  \best 3+1 & \best 2+1 & \best 3+1 & \best 1+0 & \best 1+0 & \best 1+0 & \best 5+1 & \best 3+1 \\
    \hline
\end{tabular}}
    }

    \subfloat[Winter 2019 comparison]{
        \scalebox{0.6}{\begin{tabular}{|c|l|c|c|c|c|c|c|c|c|c|c|c|c|}
    \cline{3-14}
    \multicolumn{2}{c|}{} & \multicolumn{12}{c|}{Winter 2019} \\
    \cline{3-14}
    \multicolumn{2}{c|}{} & A1 & A2 & B1 & B2 & C1 & C2 & C3 & D1 & D2 & D3 & D4 & D5 \\
    \hline
    \multirow{2}{*}{Q\#} & Lines & \best 3 & 20 & 21 & 30 & 18 & 27 & 19 & \best 3 & 12 & \best 5 & 21 & 10 \\
                         \cline{2-14}
                         & QPA & \best 1 & 8 & \best 11 & 14 & 6 & 10 & 9 & \best 1 & 8 & 5 & 11 & 9 \\
    \hline
    \multirow{2}{*}{SILQ} & lines & 10 & 10 & \best 17 & \best 15 & \best 7 & 11 & \best 7 & 4 & 15 & 18 & 17 & 15 \\
                          \cline{2-14}
                          & QPA & 4 & 8 & 12 & 12 & \best 2 & 2 & \best 2 & 2 & 5 & 7 & 7 & 8 \\
    \hline
    \multirow{2}{*}{Q-Pragma} & Lines & 6+0 & \best 8+1 & \best 17+4 & \best 12+3 & 10+1 & \best 9+1 & \best 7+1 & 4+1 & \best 7+1 & 6+1 & \best 7+1 & \best 9+0 \\
                              \cline{2-14}
                              & QPA & 4+0 & \best 2+0 & \best 10+1 & \best 8+1 & 4+0 & \best 1+0 & \best 2+0 & 2+0 & \best 3+0 & \best 4+0 & \best 4+0 & \best 7+0 \\
    \hline
\end{tabular}}
    }

    \subfloat[Mean]{
        \scalebox{0.6}{\begin{tabular}{|c|l|c|}
    \cline{3-3}
    \multicolumn{2}{c|}{} & Mean \\
    \hline
    \multirow{2}{*}{Q\#} & Lines & 16.30 \\
                         \cline{2-3}
                         & QPA & 5.59 \\
    \hline
    \multirow{2}{*}{SILQ} & lines & 9.07 \\
                          \cline{2-3}
                          & QPA & 4.81 \\
    \hline
    \multirow{2}{*}{Q-Pragma} & Lines & \best 7.26+1.67 \\
                              \cline{2-3}
                              & QPA & \best 3.15+0.33 \\
    \hline
\end{tabular}}
    }

    \caption{Evaluation of Q-Pragma compared to SILQ and Q\#, using algorithms from Q\# coding contests (Summer 2018 \cite{S18} and Winter 2019 \cite{W19}). The comparison is based on the code length, and the number of quantum primitives / annotations (QPA) used. For Q-Pragma, the code has been implemented on the Host or on the QPU: the ``+'' symbol shows the figures for both the Host and the QPU implementation (e.g. $17 + 4$ means that the Host implementation requires $17$ lines of code, while the QPU implementation requires $17+4 = 21$ lines of code). The green cells show the lowest value for each column.}
    \label{tab:bench}
\end{table}

To ensure a fair evaluation, the criteria defined in \cite{silq} (Appendix H) have been reused, and any unreadable code-shortening transformation have been avoided. The Q-Pragma primitives and annotations are:

\begin{itemize}
    \item quantum usual gates (H, X, CNOT \dots) and their derivatives (\textit{wall}, \textit{ctrl}, \textit{dag} and \textit{ctrl\_dag}),
    \item measurements and resets,
    \item pragma directives.
\end{itemize}

Quantum registers are not considered as a primitive / annotation, matching a criteria defined in \cite{silq} for Q\# (qubits are not considered as a primitive / annotation). According to this protocol, Table \ref{tab:bench} shows that, in average, Q-Pragma implementations are shorter, and use less quantum primitives / annotations than Q\# or SILQ. However, the readability of the code is key, to ensure maintainability. In the following, some implementations are provided. Q\# implementations are available in \cite{S18} and \cite{W19} while some of SILQ implementations are available on their website (\href{https://silq.ethz.ch/examples}{https://silq.ethz.ch/examples}).

\subsection{Creating a superposition (Summer 2018 - A2)}
The goal is to generate the following quantum state on $n$ qubits:
$$
\frac{1}{\sqrt{2}}\left( \ket{0} + \ket{b} \right),
$$
where $b \in \{0,1\}^n$ and $b_0 = 1$, so $b$ can also be considered as an integer such that $1 \leq b < 2^n$.

\begin{code}
\begin{lstlisting}[language=C++, caption={Implementation of $\frac{1}{\sqrt{2}}\left( \ket{0} + \ket{b} \right)$ state using Q-Pragma}, label={lst:a2}]
#include "qpragma.h"

using namespace qpragma;

|\pragma{quantum routine} (\cuint{64}bstate)|
template <|\cuint{64}|SIZE>
void solve(const |\qbool|& head,
           const |\quint{SIZE}|& tail) {
    H(head);
    
    |\pragma{quantum ctrl(head)}|
    tail += (bstate >> 1);
}
\end{lstlisting}
\end{code}

Listing \ref{lst:a2} shows a solution for this problem using Q-Pragma. This example highlights how typing registers simplifies the code development. The quantum register is splitted in two (\textit{head} and \textit{tail}). The variable $b$ being an integer, it can be added to the \textit{tail} register, as this register is a quantum integer. This idea not only shortens the code, but also makes it clearer. In this example, the first qubit is initialized with the $\ket{+}$ state, and then controls the operation updating \textit{tail} in-place: this naturally creates the superposition of the states $\ket{0}\ket{0...0}$ and $\ket{1}\ket{b_1 ... b_{n-1}}$. Because $b_0 = 1$, this routine creates the superposition $\ket{0}$ and $\ket{b}$. Note that this routine can be called on a unique quantum register, as this routine is not \textit{typed}.

\subsection{Creating a W state (Summer 2018 - A4)}
The goal is to generate a $W$ state on $N$ qubits (where $N$ is a power of $2$):
$$
  W_N = \frac{1}{\sqrt{N}}\left(\ket{100\cdots0} + \ket{010\cdots0} + \dots + \ket{000\cdots1} \right) 
$$

\begin{code}
\begin{lstlisting}[language=C++, caption={$W$ state implementation using Q-Pragma}, label={lst:a4}]
#include "qpragma.h"

using namespace qpragma;

// In this sample of code, LOG is expected
// to be equal to log2(N)
|\pragma{quantum routine}|
template <|\cuint{64}|LOG, |\cuint{64}|SIZE = (1 << LOG)>
void solve(const |\quint{SIZE}|& qreg) {
    |\quint{SIZE}|anc;
    wall::H<LOG>(anc);
    
    for (|\cuint{64}|idx = 0 ; idx < SIZE ; ++idx) {
        qreg[idx] ^= (anc == idx);

        |\pragma{quantum ctrl}| (qreg == 1 << idx)
        anc -= idx;
    }
}
\end{lstlisting}
\end{code}

In Listing \ref{lst:a4}, the creation of the $W$ state relies on an array of ancilla qubits initialized with a uniform superposition. These ancilla qubits represent all the possible indexes for the $1$ in the final $W$ state. At the end of the routine, these ancilla qubits need to be reset to state $\ket{0}$. To do so, the routine implements the following steps:

\begin{itemize}
    \item Create state $\sum \ket{i}\ket{0}$ by allocating the \textit{anc} register and applying a wall of Hadamard gate on it.
    \item Perform the transformation $\ket{i}\ket{0} \rightarrow \ket{i}\ket{2^i}$ (this is done by the first line of the \textit{for} loop).
    \item Implement the transformation $\ket{i}\ket{2^i} \rightarrow \ket{0}\ket{2^i}$ to reset the \textit{anc} register to $\ket{0}$ (done by the \textit{pragma quantum ctrl} in the \textit{for} loop).
\end{itemize}

These steps are used to create the state $\sum \ket{0}\ket{2^i}$, which correspond to the $W$ state. The two last steps can be implemented using a single \textit{for} loop (instead of $2$), shortening the implementation.

\subsection{Pattern of increasing blocks (Winter 2019 - D2)}
The goal is to implement a unitary operation which is represented by a square matrix with increasing blocks. For example, a such $3$-qubits matrix should have the following shape:

$$
\begin{pmatrix}
    * & & & & & & & \\
    & * & & & & & & \\
    & & * & * & & & &  \\
    & & * & * & & & &  \\
    & & & & * & * & * & * \\
    & & & & * & * & * & * \\
    & & & & * & * & * & * \\
    & & & & * & * & * & *
    
\end{pmatrix}
$$

\begin{code}
\begin{lstlisting}[language=C++, caption={Increasing blocks pattern implementation using Q-Pragma}, label={lst:d2}]
#include "qpragma.h"

using namespace qpragma;

|\pragma{quantum routine}|
template <|\cuint{64}|SIZE>
void solve(const std::array<|\qbool|, SIZE-1> & most,
           const |\qbool|& tail) {
    if |\constexpr|(SIZE > 1UL) {
        wall::H<SIZE - 1>.ctrl(tail, most);
        solve<SIZE - 1>.ctrl(|\castqbool|not tail, most);
    }
}
\end{lstlisting}
\end{code}

Listing \ref{lst:d2} provides an implementation of the solution given in \cite{W19}. This example relies on recursive calls of the \textit{solve} routine. The routine arguments are \textit{most} and \textit{tail}, which corresponds respectively to the whole array except for the last qubit, and the last qubit itself. When the quantum routine \textit{solve$\left<n\right>$} is called with a quantum array of $n$-qubits, this array is automatically splitted in the array \textit{most} (of size $n-1$) and the qubit \textit{tail} by Q-Pragma. Consequently, it makes the recursive call easier.

\end{document}